# How many imputations do you need?
# A two-stage calculation using a quadratic rule

Paul T. von Hippel
University of Texas, Austin

*Abstract*

When using multiple imputation, users often want to know how many imputations they need. An old answer is that 2 to 10 imputations usually suffice, but this recommendation only addresses the efficiency of point estimates. You may need more imputations if, in addition to efficient point estimates, you also want standard error (SE) estimates that would not change (much) if you imputed the data again. For replicable SE estimates, the required number of imputations increases quadratically with the fraction of missing information (not linearly, as previous studies have suggested). I recommend a two-stage procedure in which you conduct a pilot analysis using a small-to-moderate number of imputations, then use the results to calculate the number of imputations that are needed for a final analysis whose SE estimates will have the desired level of replicability. I implement the two-stage procedure using a new Stata command called **how_many_imputations** (available from [SSC](SSC)) and a new SAS macro called ***%mi_combine*** (available from the website *[missingdata.org](missingdata.org)*).

# 1  OVERVIEW

Multiple imputation (MI) is a popular method for repairing and analyzing data with missing values (Rubin, 1987). Using MI, you fill in the missing values with imputed values that are drawn at random from a posterior predictive distribution that is conditioned on the observed values $Y_{obs}$. You repeat the process of imputation $M$ times to obtain $M$ imputed datasets, which you analyze as though they were complete. You then combine the results of the $M$ analyses to obtain a point estimate $\hat{\theta}_{MI}$ and standard error estimate $\widehat{SE}_{MI}$ for each parameter $\theta$.

Because MI estimates are obtained from a random sample of $M$ imputed datasets, MI estimates include a form of random sampling variation, known as *imputation variation*. Imputation variation makes MI estimates somewhat *inefficient* in the sense that the estimates obtained from a sample of $M$ imputed datasets are more variable, and offer less power, than the asymptotic estimates that would be obtained if you could impute the data an infinite number of times. In addition, MI estimates can be *non-replicable* in the sense that the estimates you report from a sample of $M$ imputed datasets can differ substantially from the estimates that someone else would get if they re-imputed the data and obtained a different sample of $M$ imputed datasets.

Non-replicable results reduce the openness and transparency of scientific research (Freese, 2007). In addition, the possibility of changing results by re-imputing the data may tempt researchers to capitalize on chance (intentionally or not) by imputing and re-imputing the data until a desired result, such as *p*<.05, is obtained.

The problems of inefficiency and non-replicability can be reduced by increasing the number of imputations—but how many imputations are enough? Until recently, it was standard advice that $M$ = 2 to 10 imputations usually suffice (Rubin, 1987), but that advice only addresses the relative efficiency of the point estimate $\hat{\theta}_{MI}$ (see Section 4.1). If you also want the SE estimate $\widehat{SE}_{MI}$ to be replicable, then you may need more imputations—as many as $M$ > 200 in some situations—though fewer imputations may still suffice in many settings (Bodner, 2008; Graham, Olchowski, & Gilreath, 2007; White, Royston, & Wood, 2011),

It might seem conservative to set $M$ = 200 (say) as a blanket policy—and that policy may be practical in situations where the data are small and the imputations and analyses run quickly. But if the data are large, or the imputations or analyses run slowly, then you will want to use as few imputations as you can, and you need guidance regarding how many imputations are really needed for their specific data and analysis.

## 1.1  A quadratic rule

A recently proposed rule is that, to estimate SEs reliably, you should choose the number of imputations according to the linear rule $M = 100\gamma_{mis}$ (Bodner, 2008; White et al., 2011), where $\gamma_{mis}$ is the fraction of missing information, defined below in equation (4). But this linear rule understates the required number of imputations when $\gamma_{mis}$ is large (Bodner, 2008) and overstates the required number of imputations when $\gamma_{mis}$ is small (see Figure 1).

In fact, for replicable SE estimates, the required number of imputations does not fit a linear function of $\gamma_{mis}$. It is better approximated by a quadratic function of $\gamma_{mis}$. A useful version of the quadratic rule (derived in Section 4.2) is

$$M = 1 + \frac{1}{2}\left(\frac{\gamma_{mis}}{CV(\widehat{SE}_{MI}|Y_{obs})}\right)^2 \qquad (1).$$

where

$$CV(\widehat{SE}_{MI}|Y_{obs}) = \frac{SD(\widehat{SE}_{MI}|Y_{obs})}{E(\widehat{SE}_{MI}|Y_{obs})} \qquad (2).$$

is the coefficient of variation (CV), which summarizes imputation variation. The numerator and denominator of the CV are the mean $E(\widehat{SE}_{MI}|Y_{obs})$ and standard deviation $SD(\widehat{SE}_{MI}|Y_{obs})$ of the estimate $\widehat{SE}_{MI}$ across all possible sets of $M$ imputed datasets, holding the observed values $Y_{obs}$ constant.

Instead of choosing the CV directly, it may be more natural to set a target for $SD(\widehat{SE}_{MI}|Y_{obs})$. For example, if you want to ensure that the SE estimate would typically change by about .01 if the data were re-imputed, you might set a target of $SD(\widehat{SE}_{MI}|Y_{obs}) = .01$ and calculate the implied target for the CV. I will illustrate this by example in Section 2.

An alternative is to choose $M$ to achieve some desired degrees of freedom (*df*) in the SE estimate (Allison, 2003, n. 7; von Hippel, 2016, Table 3). Then, as shown in Section 4.3, the quadratic rule can be rewritten as

$$M = 1 + df\,\gamma_{mis}^2 \qquad (3)$$

Some users will want to aim for replicable SE estimates, while others may find it more natural to aim for a target *df* such as 25, 100, or 200. The two approaches are equivalent. However, estimates of *df* can be very unstable, as we will see in Section 2.8, and this makes the estimated *df* a fallible guide for the number of imputations.

## 1.2   A two-stage procedure

If you knew both $\gamma_{mis}$ and your desired CV (or your desired <u>df</u>), you could plug those numbers into formula (1) (or (13)) and get the required number of imputations $M$. But the practical problem with this approach is that $\gamma_{mis}$ is typically not known in advance. $\gamma_{mis}$ is not just the fraction of *values* that are missing; instead it is the fraction of *information* that is missing about the parameter $\theta$ (see section 4.1). $\gamma_{mis}$ is typically estimated using MI (see section 4.2), so estimating $\gamma_{mis}$ requires am initial choice of $M$.

For that reason, I recommend a two-stage procedure.

1. In the first stage, you conduct a small-$M$ pilot analysis to obtain a pilot SE estimate $\widehat{SE}_{MI}$ and a conservative estimate of $\gamma_{mis}$. From the pilot, you use the estimate $\widehat{SE}_{MI}$ and your goal for $SD(\widehat{SE}_{MI}|Y_{obs})$ to estimate your target CV as $\widehat{CV} = SD(\widehat{SE}_{MI}|Y_{obs})/\widehat{SE}_{MI}$.[1]

2. In the second stage, you plug your estimate of $\gamma_{mis}$ and your target $\widehat{CV}$ into formula (1) to get the required number of imputations $M$. If your pilot $M$ was at least as large as the required $M$, then your pilot analysis was good enough, and you can stop. Otherwise, you conduct a second and final analysis using the $M$ that was recommended by the pilot analysis.

I said that the pilot analysis should be used to get a *conservative* estimate of $\gamma_{mis}$. Let me explain what I mean by *conservative*. The obvious estimate to use from the pilot analysis is the point estimate $\hat{\gamma}_{mis}$, calculated by formula (7) below, which is output by most MI software. But this point estimate is not conservative since it runs about a 50% chance of being smaller than the true value of $\gamma_{mis}$. And if the point estimate $\hat{\gamma}_{mis}$ is too small, then the recommended number of imputations $M$ will be too small as well. So the approach runs a 50% chance of underestimating how many imputations you really need.

A more conservative approach is this: instead of using the pilot analysis to get a point estimate $\hat{\gamma}_{mis}$, use it to get a 95% confidence interval for $\gamma_{mis}$, and then use the upper bound of the confidence interval to calculate the required number of imputations $M$ from formula (1). Since there is only a 2.5% chance that the true value of $\gamma_{mis}$ exceeds the upper bound of the confidence interval, there is only a 2.5% chance that the implied number of imputations will be too low to achieve the desired degree of replicability.[2]

The two-stage procedure for deciding the number of imputations requires a CI for $\gamma_{mis}$. A CI for $logit(\gamma_{mis})$ is $logit(\hat{\gamma}_{mis}) \pm z\sqrt{2/M}$, where $z$ is a standard normal quantile (Harel, 2007).[3] This CI for $logit(\gamma_{mis})$ can be transformed to a CI for $\gamma_{mis}$ by applying the inverse-logit transformation to the endpoints. Table 1 gives 95% CIs for different values of $M$ and $\hat{\gamma}_{mis}$. Many of the confidence intervals are wide, indicating considerable uncertainty about the true value of $\gamma_{mis}$, especially when $M$ is small or $\hat{\gamma}_{mis}$ is close to .50.

In the rest of this article, I will illustrate the two-stage procedure and evaluate its properties in an applied example where about 40% of information is missing. I then test the two-stage procedure and the quadratic rule by simulation, and derive the underlying formulas. I offer a Stata command called **how_many_imputations** and a SAS macro called *%MI_COMBINE*, which recommend the number of imputations needed to achieve a desired level of replicability. The Stata command can be installed by typing **ssc install how_many_imputations** on the Stata command line. The SAS macro can be downloaded from the website *missingdata.org*.

Table 1. 95% CIs for the fraction of missing information $\gamma_{mis}$

| Point estimate $\hat{\gamma}_{mis}$ | Number of imputations $M$ | 95% CI for $\gamma_{mis}$ |
|---|---|---|
| .1 | 5 | (.03, .28) |
|    | 10 | (.04, .21) |
|    | 15 | (.05, .19) |
|    | 20 | (.06, .17) |
| .3 | 5 | (.11, .60) |
|    | 10 | (.15, .51) |
|    | 15 | (.17, .47) |
|    | 20 | (.19, .44) |
| .5 | 5 | (.22, .78) |
|    | 10 | (.29, .71) |
|    | 15 | (.33, .67) |
|    | 20 | (.35, .65) |
| .7 | 5 | (.40, .89) |
|    | 10 | (.49, .85) |
|    | 15 | (.53, .83) |
|    | 20 | (.56, .81) |
| .9 | 5 | (.72, .97) |
|    | 10 | (.79, .96) |
|    | 15 | (.81, .95) |
|    | 20 | (.83, .94) |

## 2  APPLIED EXAMPLE

This section illustrates the two-stage procedure with real data. All code and data used in this section are provided on the website *missingdata.org*.

### 2.1  Data

The SAS dataset *bmi_observed* contains data on body mass index (BMI) from the Early Childhood Longitudinal Study, Kindergarten cohort of 1998-99 (ECLS-K). The ECLS-K is a federal survey overseen by the National Center for Education Statistics, US Department of Education. The ECLS-K started with a nationally representative sample of 21,260 US kindergarteners in 1,018 schools, and took repeated measures of children's BMI in 7 different rounds, starting in the fall of 1998 (kindergarten) and ending in the spring of 2007 (8$^{th}$ grade for most students).

I estimate mean BMI in round 3 (fall of 1$^{st}$ grade). While every round of the ECLS-K missed some BMI measurements, round 3 missed the most, because in round 3 the ECLS-K saved resources by limiting BMI measurements to a random 30% subsample of participating schools. So 76% of BMI measurements were missing from round 3—70% were missing by design (completely at random), and a further 6% were missing (probably not at random) because the child was unavailable or refused to be weighed. The fraction of missing information would be 76% if all we had were the observed BMI values at round 3, but we will reduce that fraction substantially by imputation.

(The ECLS-K is a complex random sample, and in a proper analysis both the analysis model and the imputation model would account for complexities of the sample including clusters, strata, and sampling weights. But in this article, I neglect those complexities and treat the ECLS-K like a simple random sample. I did carry out an analysis that accounted for the sample's complexities. The two-stage method still performed as expected, but the estimated SEs were 50% larger and the recommended number of imputations was twice as large.)

## 2.2 Listwise deletion

The simplest estimation strategy is listwise deletion, which uses only the BMI values that are observed in round 3. The listwise estimate for mean BMI is 16.625 with an estimated SE of .037. There is little bias, because the observed values are almost a random sample of the population (over 90% of missing values are missing completely at random). But the listwise SE is larger than necessary, because it is calculated using just 24% of the sample. The results will show that MI can reduce the SE by one-third—an improvement that is equivalent to more than doubling the number of observed BMI values.

## 2.3 Multiple imputation

Next, I multiply impute missing BMIs. In each MI analysis, I get $M$ imputations of missing BMIs from a multivariate normal model for the BMIs in rounds 1-4.[4] One implication of this model is that missing BMIs from round 3 are imputed by normal linear regression on BMIs that were observed for the same child in other rounds (Schafer, 1997).[5] The imputation model predicts BMI very well; a multiple regression of round 3 BMIs on BMIs in rounds 1, 2, and 4 has $R^2=.85$, and even a simple regression of round 3 BMIs on round 2 BMIs has $R^2=.77$. The accuracy of the imputed values, and the fact that so many values are missing, ensures that the MI estimates will improve substantially over the listwise deleted estimates (von Hippel & Lynch, 2013).

I analyze each of the $M$ imputed datasets as though it were complete; that is, I estimate the (weighted) sample mean and SE in each of the $M$ samples. I then combine the $M$ to produce an MI point estimate $\hat{\theta}_{MI}$ and an SE estimate $\widehat{SE}_{MI}$. I also calculate a point estimate $\hat{\gamma}_{mis}$ and 95% confidence interval for $\gamma_{mis}$, and an estimate $\widehat{df} = (M-1)\hat{\gamma}_{mis}^{-2}$ for the $df$ of the SE estimate. Formulas for all these quantities will be given in Section 4.

## 2.4 Two-stage procedure

The two-stage procedure, which I implemented in SAS code (*two_step_example.sas*), proceeds as described earlier.

1. In the first stage, I carry out a small-$M$ pilot analysis and use the upper limit of the confidence interval for $\gamma_{mis}$ to calculate, using formula (1), how many imputations would be needed to achieve my target CV. I chose my CV goal by deciding that I wanted to ensure that the 2nd significant digit (3rd decimal place) of $\widehat{SE}_{MI}$ would typically change by about 1 if the data were re-imputed. This implies a goal of $SD(\widehat{SE}_{MI}|Y_{obs}) = .001$, so that the target CV is $\widehat{CV} = SD(\widehat{SE}_{MI}|Y_{obs})/\widehat{SE}_{MI}$, where $\widehat{SE}_{MI}$ is the pilot SE estimate.

2. In the second stage, I carry out a final analysis with the number of imputations that the first stage suggested would be needed to achieve my target CV.

## 2.5 Results with *M*=5 pilot imputations

I first tried a stage-1 pilot analysis with *M*=5 imputations. The results were an MI point estimate of 16.642 with an SE estimate of .023. The estimated fraction of missing information was .39 with a 95% CI of (.15, .69). The upper bound of the CI implied that *M*=125 imputations should be used in stage 2. In stage 2, the final analysis returned an MI point estimate of 16.650 with an SE estimate of .021. The results of both stages are summarized in the first two rows of Table 2a.

These results are not deterministic. Imputation has a random component, so if I replicate the two-stage procedure I will get different results. Table 2a gives the results of a replication (replication 2). The stage 1 pilot estimates are somewhat different than they were the first time, and the recommended number of imputations is different as well (*M*=219 rather than 125).

Although the recommended number of *imputations* changes when I repeat the two-stage procedure, the final *estimates* are quite similar. The first time I ran the two-stage procedure, our final estimate (and SE) were 16.650 (.021); the second time we ran it, our final estimate (and SE) were 16.651 (.022). The two final point estimates differ by .001, and the final SE estimates differ by .001—which is about the difference that would be expected given my goal of $SD(\widehat{SE}_{MI}|Y_{obs}) = .001$. By contrast, the two *pilot* estimates are more different; the pilot point estimates differ by .017 and the pilot SE estimates differ by .003.

The bottom of Table 2a summarizes the stage 2 estimates across 100 replications of the two-stage procedure. (The SAS code to produce the 100 replications is in *simulation.sas*.) The SD of the 100 SE estimates is .001, which is exactly what I was aiming for. That is reassuring.

Somewhat less reassuring is the tremendous variation in the number of imputations recommended by the pilot. The recommended imputations had a mean of 97 with an SD of 61. One pilot recommended as few as 4 imputations but another recommended as many as 266. The primary[6] reason for this variation is that the recommended number of imputations is a function of the pilot CI for $\gamma_{mis}$, and that CI varies substantially across replications since the pilot has only *M*=5 imputations.

## 2.6 Results with M=20 pilot imputations

One way to reduce variability in the recommended number of imputations is to use more imputations in the pilot. If the pilot uses, say, *M*=20 imputations instead of *M*=5, the pilot will yield a narrower, more replicable CI for $\gamma_{mis}$, and this will result in a more consistent recommendation for the final number of imputations.

Table 2b uses *M*=20 pilot imputation and again summarizes the stage 2 estimates from 100 replications of the two-stage procedure. Again the SD of the 100 stage 2 SE estimates is .001, which is exactly what I was aiming for. And this time the number of imputations recommended by the pilot is not as variable. The recommended imputations had a mean of 62 with an SD of 26. There is still a wide range—one pilot recommended just 22 imputations and another recommended 167—but with *M*=20 pilot imputations the range covers just one order of magnitude. The range covered two orders of magnitude when the pilot used *M*=5 imputations.

Table 2. 100 replications of the two-stage procedure. The goal is that the SD of the stage 2 SE estimates should be .001.

a. With *M*=5 pilot imputations

| Replication | Stage | Imputations | Pt. est. | SE | df | Missing information (with 95% CI) | |
|---|---|---|---|---|---|---|---|
| 1 | 1. Pilot | 5 | 16.642 | .023 | 27 | .39 | (.15, .69) |
|   | 2. Final | 125 | 16.650 | .021 | 1313 | .30 | (.25, .36) |
| 2 | 1. Pilot | 5 | 16.659 | .026 | 14 | .53 | (.25, .80) |
|   | 2. Final | 219 | 16.651 | .022 | 1876 | .34 | (.30, .38) |
| … | … | … | … | … | … | … | … |
| 100 | 1. Pilot | 5 | 16.647 | .022 | 40 | .31 | (.12, .61) |
|   | 2. Final | 89 | 16.651 | .022 | 810 | .33 | (.27, .40) |
| Stage 2 summary (across 100 replications) | Mean | 97 | 16.651 | .022 | 840 | .34 | (.26, .42) |
|   | SD | 61 | .003 | .001 | 522 | .05 | (.06, .07) |
|   | Min | 4 | 16.633 | .020 | 11 | .16 | (.06, .30) |
|   | Max | 266 | 16.656 | .026 | 2100 | .52 | (.36, .81) |

b. With *M*=20 pilot imputations

| Replication | Stage | Imputations | Pt. est. | SE | df | Missing information (with 95% CI) | |
|---|---|---|---|---|---|---|---|
| 1 | 1. Pilot | 20 | 16.654 | .021 | 217 | .30 | (.18, .44) |
|   | 2. Final | 45 | 16.652 | .022 | 451 | .31 | (.23, .41) |
| 2 | 1. Pilot | 20 | 16.658 | .020 | 371 | .23 | (.14, .35) |
|   | 2. Final | 27 | 16.647 | .022 | 208 | .35 | (.24, .48) |
| … | … | … | … | … | … | … | … |
| 100 | 1. Pilot | 20 | 16.654 | .026 | 64 | .54 | (.39, .69) |
|   | 2. Final | 167 | 16.650 | .022 | 1209 | .37 | (.32, .42) |
| Stage 2 summary (across 100 replications) | Mean | 62 | 16.652 | .022 | 535 | .34 | (.26, .43) |
|   | SD | 26 | .002 | .001 | 233 | .04 | (.04, .05) |
|   | Min | 22 | 16.645 | .021 | 118 | .25 | (.17, .33) |
|   | Max | 167 | 16.658 | .024 | 1352 | .45 | (.36, .59) |

With *M*=20 pilot imputations, the recommended number of stage 2 imputations is not just less variable—it is also lower on average. With *M*=5 pilot imputations, the average number of recommended imputations was 97; with M=*20* pilot imputations, it is just 62. When we increased the number of pilot imputations by 15, we reduced the average number of final imputations by 35. So it wouldn't pay to lowball the pilot imputations; any time we saved by using *M*=5 instead of *M*=20 imputations in stage 1 gets clawed back double in stage 2.

Why does the recommended number of imputations rise if we reduce the number of pilot imputations? With fewer pilot imputations we are more likely to get a high pilot estimate of $\gamma_{mis}$ and this leads to a high recommendation for the final number of imputations. With fewer pilot imputations, we are also more likely to get a *low* pilot estimate of $\gamma_{mis}$, but that doesn't matter as much. Because the recommended number of imputations increases with the square $\gamma_{mis}^2$, the recommended number of imputations is more sensitive to small changes in the estimate $\gamma_{mis}$ of when that estimate is high than when it is low.

## 2.7  How many *pilot* imputations do you need?

So how many pilot estimates are enough? In a sense, it doesn't matter how many pilot imputations you use in stage 1, since the procedure almost always ensures that in stage 2 you will use enough imputations to produce SE estimates with the desired level of replicability.

In another sense, though, there are costs to lowballing the pilot imputations. If you don't use many imputations in the pilot, the number of imputations in stage 2 may be unnecessarily high on average, and unnecessarily variable. This is a particular danger when the true fraction of missing information is high, as it is in Table 2, where the fraction of missing information is about 38%.[7]

Perhaps the best guidance is to use more pilot imputations when the true value of $\gamma_{mis}$ seems likely to be large. You won't have a formal estimate of $\gamma_{mis}$ until after the pilot, but you often have a reasonable hunch whether $\gamma_{mis}$ is likely to be large or small. In my ECLS-K example, with 76% of values missing, it seemed obvious that $\gamma_{mis}$ was going to be large. I didn't know exactly how large until the results were in, but I could have guessed that *M*=5 imputations wouldn't be enough. *M*=20 pilot imputations was a more reasonable choice, and it led to a more limited range of recommendations for the number of imputations in stage 2.

If a low fraction of missing information is expected, on the other hand, then fewer pilot imputations are needed. You can use just a few pilot imputations, and they may be sufficient. If they are not, then the recommended number of imputations in stage 2 will not vary that much.

## 2.8  Why the estimated *df* is an unreliable guide

In the Overview we mentioned that the number of imputations can be chosen to ensure that the true *df* exceeds some threshold, such as 100. This is correct, but basing the number of imputations on the estimated *df* is a risky business. For example, when I used *M*=5 pilot imputations, about a quarter of my pilot analyses had an estimated *df* > 100. So in about a quarter of pilot analyses, I would have concluded that *M*=5 imputations was enough—when clearly it is not enough for an SE with the desired level of replicability. So the estimated *df* is an unreliable guide to whether you have enough imputations.

To understand this instability, it is important to distinguish between the true *df* and the estimated *df*. The true *df* is $df = (M-1)\gamma_{mis}^{-2}$, but the estimate is $\widehat{df} = (M-1)\hat{\gamma}_{mis}^{-2}$ and this estimate is very sensitive to estimation error in $\hat{\gamma}_{mis}$. So the fact that the estimated *df* exceeds 100 is no guarantee that the true *df* exceeds 100—and no guarantee that you have enough imputations.

# 3 VERIFYING THE QUADRATIC RULE

The two-stage procedure rests on the quadratic rule (1), so it is important to verify that the rule is approximately correct. In Section 4, we will derive the quadratic rule analytically. Here we verify that it approximately fits the results of a simulation published by Bodner (2008).

In his simulation, Bodner varied $\gamma_{mis}$ and estimated how many imputations $M$ were needed to satisfy a criterion very similar to $CV(\widehat{SE}_{MI}|Y_{obs}) = .05$. Figure 1 fits Bodner's results (2008, Table 3, column 2) to our quadratic rule which, with a CV of .05, simplifies to $M = 200\gamma_{mis}^2$. Figure 1 also shows the linear rule $M = 100\gamma_{mis}$ proposed by others (Bodner, 2008; White et al., 2011).

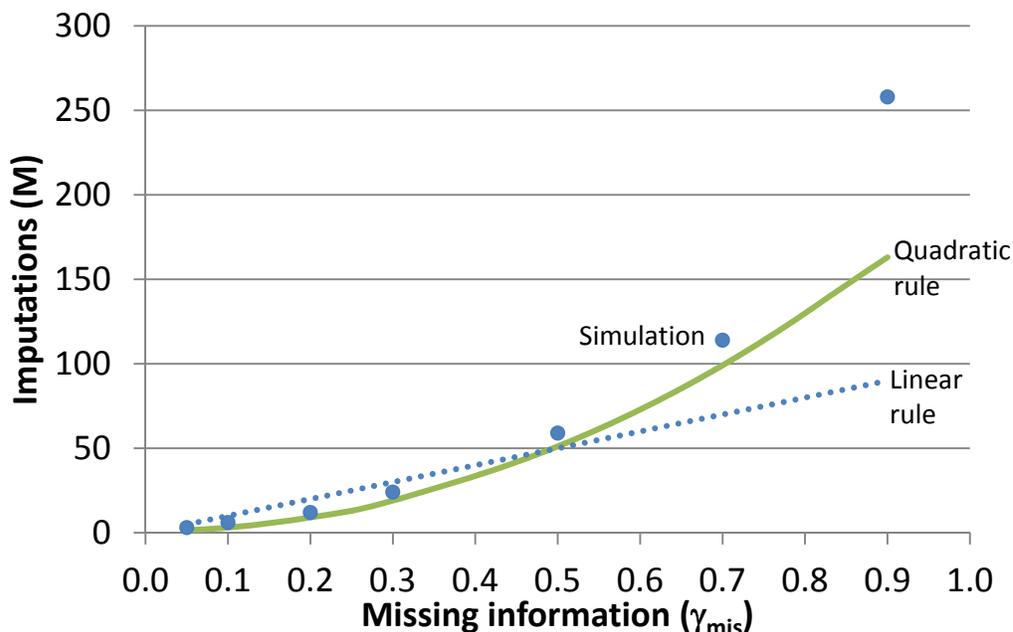

*Figure 1.* Results of Bodner's (2008) simulation fitted to the linear rule $M = 100\gamma_{mis}$ and to the quadratic rule, which in this situation simplifies to $M = 200\gamma_{mis}^2$.

Clearly the quadratic rule fits better. The two rules agree at $\gamma_{mis} = .5$, but the linear rule somewhat overstates the number of imputations needed when $\gamma_{mis} < .5$, and substantially understates the number of imputations needed when $\gamma_{mis} > .5$.

While the quadratic rule fits better, it does slightly under-predict the number of imputations that were needed in Bodner's simulation. Possible reasons for this include the fact that Bodner's criterion was not exactly $CV(\widehat{SE}_{MI}|Y_{obs}) = .05$. In particular,

- Bodner's criterion did not pertain directly to $\widehat{SE}_{MI}$. Instead, it pertained to the half-width $\hat{h}_{MI} = t\widehat{SE}_{MI}$ of a 95% confidence interval $\hat{\theta}_{MI} \pm t\widehat{SE}_{MI}$, where $t$ is the .975 quantile from a t distribution with $\widehat{df} = (M-1)\hat{\gamma}_{mis}$. When $\gamma_{mis}$ is small, $t$ is practically a constant since $\widehat{df}$ is large—so managing the variation in $\widehat{SE}_{MI}$ is equivalent to the variation in $\hat{h}_{MI}$. But when $\gamma_{mis}$ is large, $t$ contains imputation variation since $\widehat{df}$ may be small, and more imputations may be needed to manage the variation in $t$ along with the variation in $\widehat{SE}_{MI}$. In other words, the quadratic rule may suffice to manage the variation in $\widehat{SE}_{MI}$, but more imputations may be needed to manage the variation in $\hat{h}_{MI}$, especially when $\gamma_{mis}$ is large.

- Bodner's criterion also did not pertain directly to the CV of $\hat{h}_{MI}$. Instead, his criterion was that, across many sets of $M$ imputations, 95% of $\hat{h}_{MI}$ values would be within 10% of the mean $E(\hat{h}_{MI})$. This is equivalent to the criterion $CV(\hat{h}_{MI}|Y_{obs}) = .05$, as $M$ grows and the distribution of $\hat{h}_{MI}$ approaches normality. But the criteria are not equivalent if $M$ is smaller.

Another consideration is that our expression for $CV(\widehat{SE}_{MI}|Y_{obs})$ is approximate rather than exact. We derive our expression in Section 4.

## 4 FORMULAS

In this section, we derive formulas for the number of imputations that are required for different purposes. Some of these formulas were given in the Overview, and are now justified. Other formulas will be new to this section.

The number of imputations that is required depends on the quantity that is being estimated. Relatively few imputations are needed for an efficient point estimate $\hat{\theta}_{MI}$. More imputations are needed for a replicable SE estimate, and even more imputations may be needed for a precise estimate of $\gamma_{mis}$.

### 4.1 Point estimates

Suppose you have a sample of $N$ cases, some of which are incomplete. MI makes make $M$ copies of the incomplete dataset. In the $m^{th}$ copy, MI fills in the missing values with random imputations from the posterior predictive distribution of the missing values given the observed values (Rubin, 1987).

You analyze each of the $M$ imputed dataset as though it were complete and obtain $M$ point estimates $\hat{\theta}_m$, $m=1,\ldots,M$ for some parameter of interest $\theta$, such a mean or a vector of regression parameters. You average the $M$ point estimates to get an MI point estimate $\hat{\theta}_{MI} = \sum_{m=1}^{M} \hat{\theta}_m/M$.

The true variance of the MI point estimate is $V_{MI} = V(\hat{\theta}_{MI})$. If $\hat{\theta}_{MI}$ is a scalar, then $V_{MI} = SE_{MI}^2$ is the square of its SE. If $\hat{\theta}_{MI}$ is a vector, then each component has its own value of $V_{MI}$.[8]

$V_{MI}$ reflects two sources of variation: *sampling variation* due to the fact that we could have taken a different sample of $N$ cases, and *imputation variation* due to the fact that we could have taken a different sample of $M$ imputations. You can reduce imputation variation by increasing the number of imputations $M$ so that $V_{MI}$ converges toward a limit that reflects only sampling variation. This is the infinite-imputation variance: $V_{\infty I} = \lim_{M \to \infty} V_{MI}$. (Throughout this Section, the subscript $\infty I$ will be used for the limit of an MI expression as $M \to \infty$.)

Although $V_{\infty I}$ is the lower bound for the variance you can achieve by applying MI to the incomplete data, $V_{\infty I}$ is still greater than the variance that you could have achieved if the data were complete: $V_{com}$. The ratio $\gamma_{obs} = V_{com}/V_{MI}$ is known as the *fraction of observed information*, and its complement is the *fraction of missing information* $\gamma_{mis}$:

$$\gamma_{mis} = 1 - V_{com}/V_{MI} \qquad (4)$$

Note that the fraction of missing information is generally *not* the same as the fraction of values that are missing. Typically the fraction of missing information is smaller than the fraction of missing values, but one can contrive situations where it is larger.

Since you have missing values, you are not going to achieve the complete-data variance $V_{com}$, and since you can't draw infinite imputations you are not even going to achieve the infinite-imputation variance $V_{\infty I}$. But how many imputations do you need to come reasonably close to $V_{\infty I}$? The traditional advice is that $M = 2$ to $10$ imputations are typically enough. The justification for this is the following formula (Rubin, 1987)

$$V_{MI} = V_{\infty I}(1 + \gamma_{mis}/M) \qquad (5),$$

which says that the variance of an MI point estimate with $M$ imputations is only $(1 + \gamma_{mis}/M)$ times larger than it would be with infinite imputations. For example, even with $\gamma_{mis}=80\%$ missing information, choosing $M = 10$ means that the variance of $\hat{\theta}_{MI}$ is just 8% larger (so the SE is only 4% larger) than it would be with infinite imputations. In other words, increasing $M$ beyond 10 can only reduce the SE by less than 4%.

## 4.2 Variance estimates

The old recommendation of $M = 2$ to $10$ is fine if all you want is an efficient point estimate $\hat{\theta}_{MI}$. But what you also want an efficient and replicable estimate $\hat{V}_{MI}$ of the variance $V_{MI}$, or equivalently an efficient and replicable estimate $\widehat{SE}_{MI}$ for the standard error $SE_{MI} = V_{MI}^{1/2}$? Then you may need more imputations—perhaps considerably more.

The most commonly used variance estimator for multiple imputation is

$$\hat{V}_{MI} = \hat{W}_{MI} + \left(1 + \frac{1}{M}\right)\hat{B}_{MI} \qquad (6)$$

(Rubin, 1987). Here $\hat{B}_{MI} = \sum_{m=1}^{M}(\hat{\theta}_m - \hat{\theta}_{MI})^2/(M-1)$ is the *between variance*—the variance of the point estimates across the $M$ imputed datasets. $\hat{W}_{MI} = \sum_{m=1}^{M}\hat{W}_m/M$ is the *within variance*—the average of the $M$ squared SEs ($\hat{W}_m = \widehat{SE}_m^2$) each estimated by analyzing an imputed dataset as though it were complete.

The within variance $\hat{W}_{MI}$ consistently estimates $V_{com}$ which again is the variance you would get if the data were complete. Therefore the fraction of observed information $\gamma_{obs} = V_{MI}^{-1}V_{com}$ can be estimated by the ratio $\hat{\gamma}_{obs} = \hat{V}_{MI}^{-1}\hat{W}_{MI}$, and the fraction of missing information $\gamma_{mis} = 1 - \gamma_{obs}$ is estimated by the complement (Rubin, 1987):

$$\hat{\gamma}_{mis} = 1 - \hat{\gamma}_{obs} \quad (7)$$
$$= \hat{V}_{MI}^{-1}\hat{B}_{MI}(1 + 1/M)$$

Notice the implication that $\hat{B}_{MI}$ tends to be larger (relative to $\hat{V}_{MI}$) if the true value of $\gamma_{mis}$ is large.

In fact, $\hat{B}_{MI}$ is the Achilles heel of the variance estimator $\hat{V}_{MI}$. Since $\hat{B}_{MI}$ is a variance estimated from a sample of size $M$, if $M$ is small $\hat{B}_{MI}$ will be *volatile* in the sense that a noticeably different value of $\hat{B}_{MI}$ may be obtained if the data are imputed again. This may not be a problem if $\gamma_{mis}$ is small, but if $\gamma_{mis}$ is large then $\hat{B}_{MI}$ will constitute a large fraction of $\hat{V}_{MI}$, so that $\hat{V}_{MI}$ will be volatile as well.

The volatility or imputation variation in $\hat{V}_{MI}$ can be summarized by the coefficient of variation $CV(\hat{V}_{MI}|Y_{obs})$. A simple approximation for the CV is

$$CV(\hat{V}_{MI}|Y_{obs}) \xrightarrow[M\to\infty]{} \gamma_{mis}\sqrt{2/(M-1)} \quad (8)$$

(von Hippel, 2007, Appendix A). Then solving for $M$ yields the following quadratic rule for choosing the number of imputations to achieve a particular value of $CV(\hat{V}_{MI}|Y_{obs})$:

$$M \approx 1 + 2\left(\frac{\gamma_{mis}}{CV(\hat{V}_{MI}|Y_{obs})}\right)^2 \quad (9)$$

This quadratic rule is uses the CV of $\hat{V}_{MI}$, but in practice you will likely be more interested in the CV of $\widehat{SE}_{MI}$. The two are related by the approximation $CV(\hat{V}_{MI}) \approx 2\,CV(\widehat{SE}_{MI})$, which follows from the delta method. And plugging this approximation into equation (10), we get

$$M \approx 1 + \frac{1}{2}\left(\frac{\gamma_{mis}}{CV(\widehat{SE}_{MI})}\right)^2 \quad (10)$$

which is the formula (1) that we gave earlier, in the Overview.

### 4.3 Degrees of freedom

Allison (2003, n. 7) recommends choosing $M$ to achieve some target degrees of freedom (*df*). This turns out to be equivalent to our suggestion of choosing $M$ to reduce the variability in $\hat{V}_{MI}$ and $\widehat{SE}_{MI}$. In large samples, when conditioned on $Y_{obs}$, $\hat{V}_{MI}$ follows approximately a chi-square distribution with $df = (M-1)\gamma_{mis}^{-2}$ (Rubin, 1987).[9] The CV of such a distribution is $CV(\hat{V}_{MI}|Y_{obs}) = \sqrt{2/df}$ which implies the relationship

$$df = \frac{2}{\left(CV(\hat{V}_{MI}|Y_{obs})\right)^2} = \frac{1}{2\left(CV(\widehat{SE}_{MI}|Y_{obs})\right)^2} \quad (11)$$

So aiming for an SE with $CV(\widehat{SE}_{MI}|Y_{obs}) = .05$, for example, is equivalent to aiming for an SE with $df=200$. Figure 2 graphs the relationship between $df$ and $CV(\widehat{SE}_{MI}|Y_{obs})$.

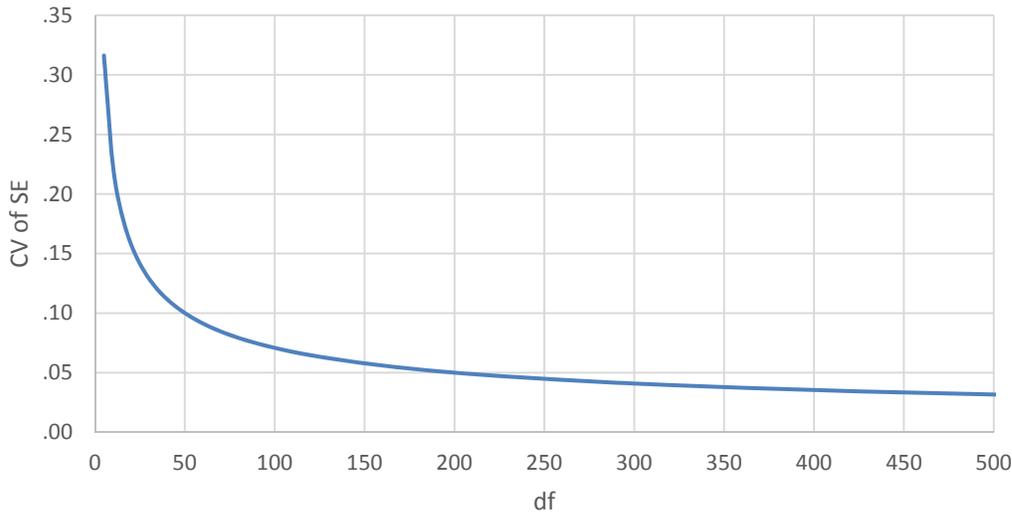

*Figure 2.* Relationship between the degrees of freedom and the coefficient of variation for an MI SE estimate.

If you are aiming for a particular $df$, you can reach it by choosing $M$ according to the following quadratic rule:

$$M = 1 + df\, \gamma_{mis}^2 \qquad (12)$$

which we presented earlier as (3). It is equivalent to the earlier quadratic rule (10), except that rule (10) was written in terms of the CV while rule (13) is written in terms of the $df$. Rule (13) can also be derived more directly, from the definition $df = (M-1)\gamma_{mis}^{-2}$.

Again, since $\gamma_{mis}$ is unknown, it makes sense to proceed in two stages. First, carry out a pilot analysis to obtain a conservatively large estimate of $\gamma_{mis}$, such as the upper limit of a 95% confidence interval. Then use that estimate of $\gamma_{mis}$ to choose a conservatively large number of imputations $M$ which will with high probability achieve the desired $df$.

Remember that the exact $df$ are unknown since they are a function of the unknown $\gamma_{mis}$. The $df$ must be estimated by $\widehat{df} = (M-1)/\hat{\gamma}_{mis}$. While the two-stage procedure can ensure with high probability that the true $df$ achieve the target, the estimated $df$ may be well off. So the estimated $df$ is an unreliable indicator of whether you have enough imputations. We saw that in Section 2.8.

## 5  CONCLUSION

How many imputations do you need? An old rule of thumb was that $M=2$ to 10 imputations is usually enough. But that rule only addressed the variability of point estimates. More recent rules also addressed the variability of SE estimates, but those rules were limited in two ways. First, they modeled the variability of SE estimates with a linear function of the fraction of missing information $\gamma_{mis}$. Second, they required the value of $\gamma_{mis}$, which is not known in advance.

I have proposed a new rule that relies on a more accurate quadratic function in which the number of imputations needed increases with the square of $\gamma_{mis}$. And since $\gamma_{mis}$ is unknown, I have proposed a two-stage procedure in which $\gamma_{mis}$ is estimated from a small-*M* pilot analysis, which serves as a guide for how many imputations to use in stage 2. The stage 1 estimate is the top of a 95% CI for $\gamma_{mis}$, which is conservative in that it ensures that we are unlikely to use too few imputations in stage 2.

To make this procedure convenient, I have written software for Stata and SAS. For Stata, I wrote the command **how_many_imputations**. On the Stata command line, install it by typing **ssc install how_many_imputations**; then type **help how_many_imputations** to learn how to use it. For SAS, I wrote the *%MI_COMBINE* macro, which is available on the website *missingdata.org*. The website also provides code to illustrate the use of the macro (*two_step_example.sas*) and to replicate all the results in this article (*simulation.sas*).

# ENDNOTES

[1] Here $\widehat{SE}_{MI}$ is a plug-in estimate for $E(\widehat{SE}_{MI}|Y_{obs})$.

[2] This two-stage procedure for choosing the number of imputations *M* is inspired by a two-stage procedure for choosing the sample size *N* for a survey or experiment. The sample size *N* required to achieve a given precision or power typically depends on parameters such as the residual variance $\sigma^2$. These parameters are typically not known in advance, but must be estimated from a small pilot sample. Because the pilot estimates may be imprecise, a common recommendation is not to calculate *N* from the pilot point estimate $\hat{\sigma}^2$, but from the high end of a confidence interval for $\sigma^2$ calculated from the pilot data (e.g., Dean & Voss, 2000; Levy & Lemeshow, 2009).

[3] The derivation of this CI is asymptotic, but in simulations it maintained close to nominal coverage in small samples with few imputations and large fractions of missing information—e.g., *n*=100 cases with 40% missing information and *M*=5 imputations (Harel, 2007, Table 2).

[4] Adding rounds 5-7 to the model increases runtime but does little to improve the $R^2$ of the imputation model.

[5] BMI does not follow a normal distribution, but it doesn't need to for a normal model to produce approximately unbiased estimates of the mean (von Hippel, 2013). Furthermore, the imputed BMIs do not follow a normal distribution. They are skewed because they are imputed by regression on observed BMIs that are skewed. Although the imputed residuals are normal, the residuals only account for 15-23% of the variance in the imputed BMI values.

[6] A secondary reason is variation in the estimate $\widehat{SE}_{MI}$ that is used to estimate the target $\widehat{CV} = SD(\widehat{SE}_{MI}|Y_{obs})/\widehat{SE}_{MI}$.

[7] I got this estimate by averaging $\hat{\gamma}_{mis}$ across stage 2 replications.

[8] Alternatively, if $\hat{\theta}_{MI}$ is a vector we can define $V_{MI}$ as a covariance matrix whose diagonal contains squared standard errors for each component of $\hat{\theta}_{MI}$. But the matrix definition is not helpful here since we don't plan to use the off-diagonal elements.

[9] This is a large-sample *df* formula. Some MI software can also output a small-sample *df* estimate that is limited by the number of observations as well as the number of imputations (Barnard & Rubin, 1999). This small-sample formula is useful for making appropriately calibrated inferences about the parameters $\theta$, but it should not be used to choose the number of imputations *M*.